\documentclass[prl,twocolumn,showpacs,superscriptaddress]{revtex4-1}
\usepackage{graphicx}
\usepackage{amssymb}
\usepackage{amsmath}

\begin{document}
\title{Influence of Magnetic Scattering on Superfluidity of $^3$He in Nematic Aerogel}
\author{V.\,V.\,Dmitriev}
\email{dmitriev@kapitza.ras.ru}
\affiliation{P.L.~Kapitza Institute for Physical Problems of RAS, 119334 Moscow, Russia}
\author{A.\,A.\,Soldatov}
\affiliation{P.L.~Kapitza Institute for Physical Problems of RAS, 119334 Moscow, Russia}
\affiliation{Moscow Institute of Physics and Technology, 141700 Dolgoprudny, Russia}
\author{A.\,N.\,Yudin}
\affiliation{P.L.~Kapitza Institute for Physical Problems of RAS, 119334 Moscow, Russia}

\date{\today}

\begin{abstract}
We report results of experiments with superfluid $^3$He confined in aerogels with parallel strands which lead to anisotropic scattering of $^3$He quasiparticles. We vary boundary conditions for the scattering by covering the strands by different numbers of atomic $^4$He layers and observe that the superfluid phase diagram and the nature of superfluid phases strongly depend on the coverage. We assume that the main reason of these phenomena is a magnetic channel of the scattering which becomes important at low coverages. Our results show that the magnetic channel also may be important in other Fermi systems with the triplet pairing.
\end{abstract}

\maketitle
{\it Introduction.---}In many Fermi systems, e.g. in liquid $^3$He, in some cold atomic gases, and unconventional superconductors, a triplet Cooper pairing occurs that results in superfluid (superconducting) states. An ideal object to study an effect of impurities on such states is $^3$He: it has a spherical Fermi surface, its superfluid phases (A, B and A$_1$) are well understood, and its superfluid coherence length can be varied by pressure in range of 20--80 nm \cite{VW}. Although  superfluid $^3$He is originally pure, a well defined system of impurities can be introduced by a high porosity aerogel. In such experiments silica aerogels are typically used. The main effect of the aerogel is to scatter $^3$He quasiparticles. At temperatures $T\sim1$\,mK, where $^3$He is superfluid, the scattering occurs only on aerogel strands and boundary conditions may be varied by a small amount of $^4$He which covers the strands by a few atomic layers. In pure $^3$He the strands are covered by $\sim2$ atomic layers of paramagnetic solid $^3$He \cite{Sch87} and the scattering is diffusive but in presence of more than $\approx2.5$ layers of $^4$He it is nearly specular at low pressures and becomes purely diffusive above $\approx25$\,bar \cite{Rich,P91,K93,Steel,M12}. The $^4$He coverage also removes the solid $^3$He, and spin is conserved during the scattering. In contrast, in pure $^3$He spin is not conserved due to a fast exchange between atoms of liquid and solid $^3$He that should result in an additional spin-exchange (magnetic) scattering channel. However, experiments with silica aerogels do not show a clear evidence of the magnetic channel. In particular, the observed A-like and B-like phases correspond to A and B phases of bulk $^3$He, regardless of presence or absence of $^4$He \cite{Osh98,Osh00,Halp02,weB,AB,we0,Halp11}. Moreover, the superfluid transition temperature of $^3$He in aerogel ($T_{ca}$) is independent on the coverage at high pressures \cite{Osh98,AB} but slightly higher in presence of the coverage at lower pressures \cite{Halp96,Parp98} probably due to the change of the scattering specularity. In most of theoretical models of $^3$He in aerogel the magnetic channel is neglected except only a few papers where it was shown that the magnetic scattering may affect A-A$_1$ transition in high magnetic fields \cite{Bar,S2003,Bar2} and a heat transport in the normal phase \cite{Sauls2010}.

Presumably, the magnetic scattering in $^3$He in silica aerogels is masked due to their small global anisotropy.
In this case the scattering is nearly isotropic, regardless of whether it is diffusive or specular, and an additional ``randomization'' due to the magnetic channel does not change the picture too much. The situation may be different for  essentially anisotropic scattering but it has not been investigated earlier. To obtain the anisotropic scattering a ``nematic'' aerogel (N-aerogel) \cite{Ask2} can be used. Its strands are nearly parallel to one another, and at ultralow temperatures an effective mean free path of $^3$He quasiparticles along the strands direction $\hat\zeta$ is longer than in the transverse direction \cite{diff0,diff}. Theoretically, it makes favorable new phases not existing in bulk $^3$He, i.e. polar, polar distorted A (denoted as DA), and polar distorted B (denoted as DB) phases \cite{AI,Sauls,Fom,Ik}. These phases were recently observed in experiments with $^3$He in N-aerogels \cite{we1,we4,we3,P1}. We note that aerogel strands in Refs.~\cite{we1,we4,we3} were covered by $\approx2.5$ atomic layers of $^4$He to remove the paramagnetic $^3$He. In this work we describe experiments with $^3$He in N-aerogels in pure $^3$He and with strands covered by 2.2 and 2.5 atomic layers of $^4$He (2.2-coverage and 2.5-coverage). We have observed that even a small amount of paramagnetic $^3$He on the strands drastically changes the superfluid phase diagram.

{\it Samples and methods.---}We used 4 samples of N-aerogel with different porosities. They have a cuboid shape with sizes of $\approx4$\,mm and were prepared from a new material ``nafen'' (produced by ANF Technology) which consists of Al$_2$O$_3$ strands with diameters $\approx9$\,nm. More systematic investigations were done with 2 samples: ``nafen-72'' cut from nafen with overall density 72\,mg/cm$^3$ (the porosity is 98.2\%, the mean distance between strands is $\approx60$\,nm) and ``nafen-910'' prepared from the same piece of nafen as described in Ref.~\cite{iet}. It has density 910\,mg/cm$^3$, porosity 78\%, so the distance between strands is only $\approx10$\,nm. Experiments in pure $^3$He and with 2.5-coverage were done also with 2 another samples used in experiments \cite{we3}: ``nafen-90'' and ``nafen-243'' with densities 90 and 243\,mg/cm$^3$ (porosities are 97.8\% and 93.9\%). The necessary temperatures were obtained by a nuclear demagnetization cryostat and determined by a quartz tuning fork. To obtain the desirable coverage we add a fixed amount of $^4$He into the empty chamber at $T<100$\,mK. We assign 2.5-coverage to the minimal amount of $^4$He which at 29.3\,bar results in absence of the paramagnetic NMR signal from solid $^3$He.

Measurements were done by continuous wave (cw) NMR in magnetic fields 2.4--27.8\,mT (NMR frequencies are 78--902\,kHz) and at pressures 0.2--29.3\,bar. An external magnetic field $\bf H$ could be oriented at any angle $\mu$ with respect to $\hat\zeta$. Similar to experiments \cite{we1,we4,we3}, superfluid phases were identified by specific NMR properties which depend on the order parameter, its spatial distribution, and $\mu$. One of these properties is a value of spin susceptibility. Polar, DA, and A phases are equal spin pairing (ESP) phases in which the susceptibility equals to that in the normal phase. DB and B phases are not the ESP states, and their susceptibility is smaller. Below we describe some other properties of polar, DA, and A phases which are important for their identification.

General form of the order parameter of these phases is
\begin{equation}
\label{A}
A_{{\nu}k} =\Delta_0 e^{i\varphi}d_{\nu}\left(am_k+ibn_k\right),
\end{equation}
where $\Delta_0$ is the gap parameter, $\varphi$ is the phase, $\bf d$ is the unit spin vector, $\bf m$ and $\bf n$ are mutually orthogonal unit orbital vectors, and $a^2+b^2=1$. The DA phase ($a^2>b^2>0$) is an intermediate state between polar ($a=1, b=0$) and A phases ($a=b$) but the polar phase is topologically different: it is not chiral and its gap is zero in the plane normal to {\bf m} \cite{AI}. Besides, strands of N-aerogel destroy a long-range order in DA and A phases: ${\bf m}\parallel\hat\zeta$ but vectors $\bf n$ are random at distances larger than $\sim1$\,$\mu$m forming a static 2D Larkin-Imry-Ma state in the plane normal to $\hat\zeta$ \cite{we1,Vol,we5}. Vectors $\bf d$ are normal to the magnetization, their uniform distribution is favorable (spin nematic, SN, state), and it is this state that is usually observed. However, in DA and A phases a metastable spin glass (SG) state with random $\bf d$ also may exist that corresponds to a local minimum of energy. The SG state can be created in NMR experiments by cooling through $T_{ca}$ with high excitation that generates a random $\bf d$-distribution \cite{we0}. On further cooling, this state is stabilized by the random $\bf n$-field due to a dipole interaction $U_D\propto\left(a^2({\bf dm})^2+b^2({\bf dn})^2\right)$. There is no $\bf n$ in the polar phase, so the SG state is unstable. Above mentioned properties  are manifested in NMR properties \cite{we3}. In the SN state a cw NMR frequency shift ($\Delta\omega$) from the Larmor frequency ($\omega_L=\gamma H$) in polar, DA, and A phases is zero for $\mu=90^\circ$. But for $\mu=0$ there is a quantitative difference, and the shift equals
\begin{equation}
\label{shift0}
2\omega_L\Delta\omega=K\Omega_A^2,
\end{equation}
where $K$ depends on $a$ and $b$, $\Omega_A$ is the Leggett frequency of the A phase (if this phase had the same transition temperature) which for small suppressions of the superfluid transition temperature (i.e. for $\Delta T_{ca}=T_c-T_{ca}\ll T_c$, where $T_c$ is the transition temperature in bulk $^3$He) can be determined from $\Omega_A$ in bulk $^3$He. In the A phase $K=1/2$ but in the polar phase, in a weak-coupling limit, $K$ should equal 4/3. As it was found in experiments \cite{we3}, in the polar phase $K$ decreases from 4/3 to 1.15 when increasing pressure from 2.9 to 29.3\,bar (presumably due to strong-coupling effects), but practically is independent from nafen porosity. Thus, measurements of $\Delta\omega$ at $\mu=0$ allow to distinguish between the phases, at least if $\Delta T_{ca}\ll T_c$. One more property is that in the SG state $\Delta\omega$ is negative for $\mu=90^\circ$. The SG state cannot be formed in the polar phase, i.e. the negative shift for $\mu=90^\circ$ means that the phase is not polar.

\begin{figure}[t]
\centerline{\includegraphics[width=\columnwidth]{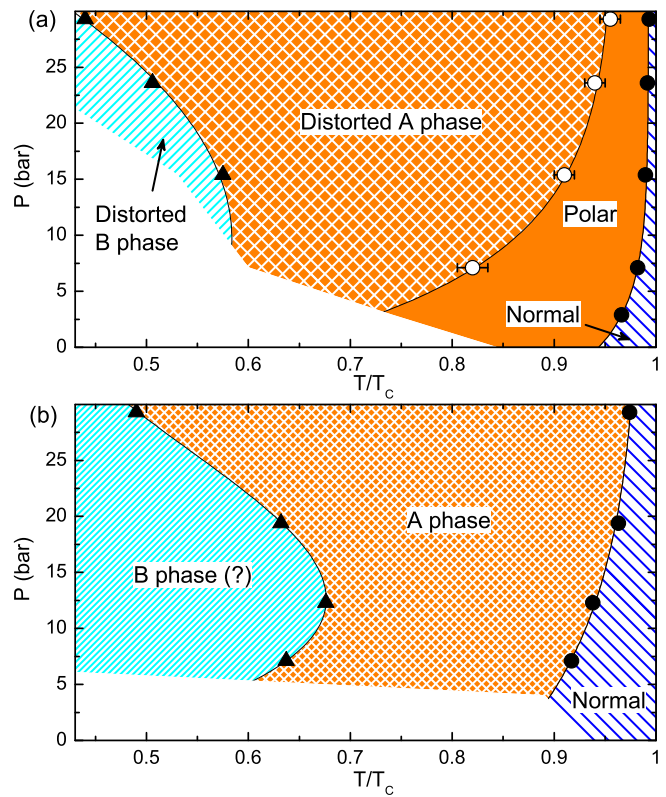}}
\caption{
Phase diagrams of $^3$He in nafen-72 for 2.5-coverage (a) and in pure $^3$He (b) obtained on cooling from the normal phase. Filled circles mark $T_{ca}$. Open circles mark the transition between polar and DA phases. Triangles mark the beginning of the transition into the DB (or B) phase. The white area shows regions with no experimental data. The x-axis represents the temperature normalized to the superfluid transition temperature in bulk $^3$He.}
\label{pd72}
\end{figure}

\begin{figure*}[t]
\includegraphics[width=\textwidth]{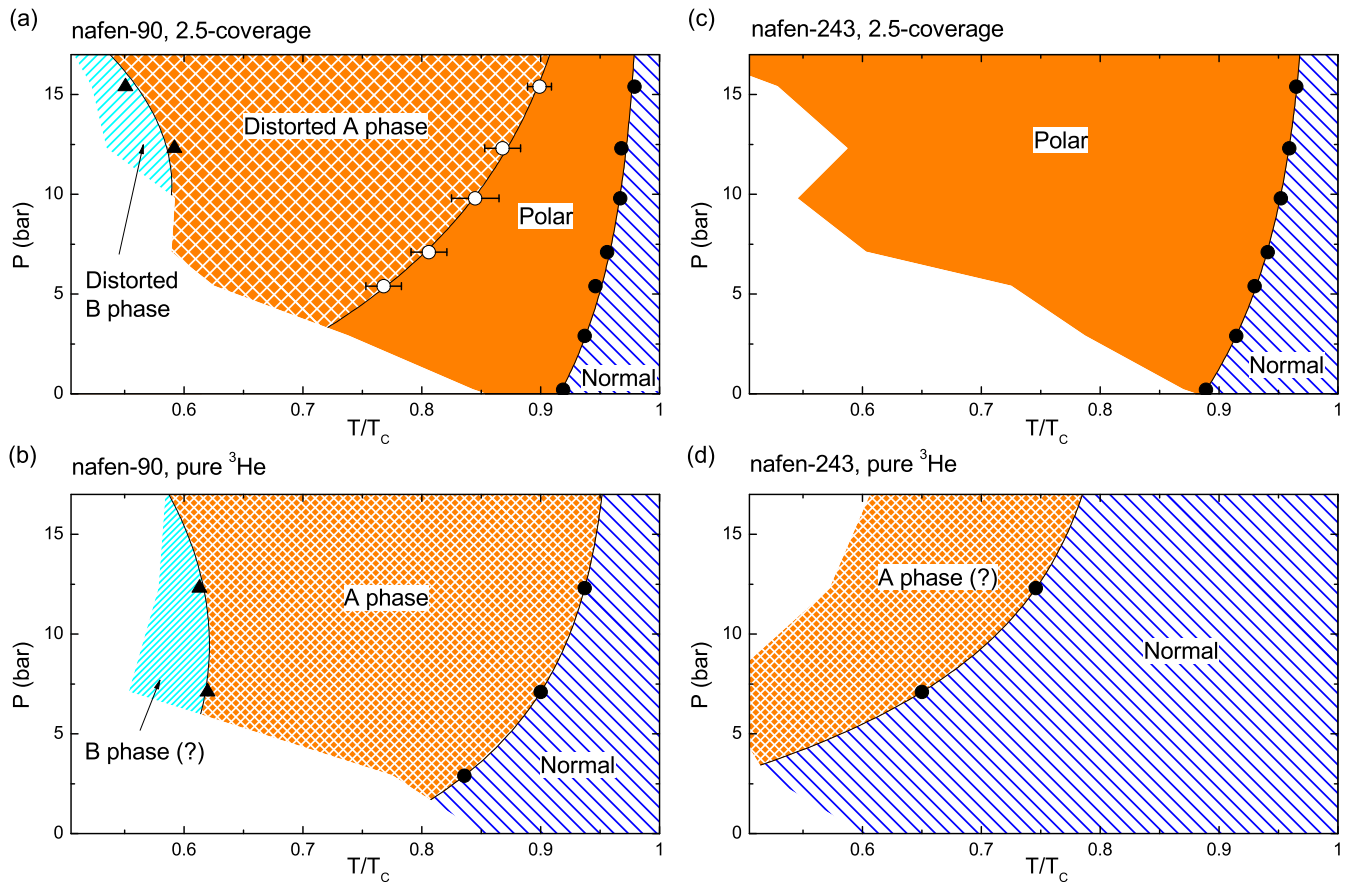}
\caption{Phase diagrams of $^3$He confined in nafen-90 (a,b) and nafen-243 (c,d) for 2.5-coverage (a,c) and in pure $^3$He (b,d). The data are obtained on cooling from the normal phase. Filled circles mark $T_{ca}$. Open circles mark the transition between polar and DA phases. Triangles mark the beginning of the transition into the DB (or B) phase. In the case of pure $^3$He in nafen-243 we are not able to distinguish whether the observed superfluid phase is the A phase or the DA phase because a rather large suppression of $T_{ca}$ does not allow us to use the rescaled bulk value of $\Omega_A$ in Eq.~\ref{shift0} for determination of $K$.}
\label{mode}
\end{figure*}

\begin{figure}[t]
\centerline{\includegraphics[width=\columnwidth]{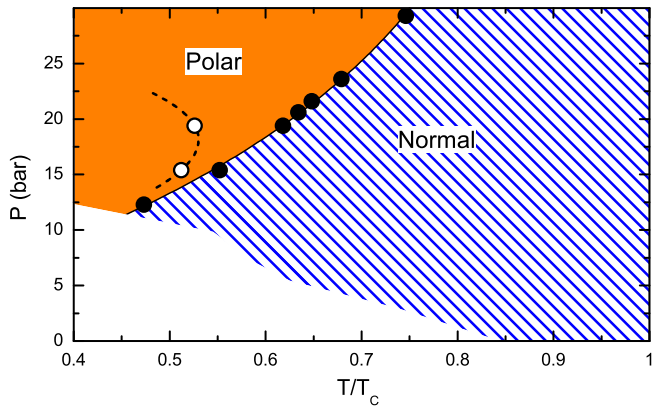}}
\caption{
Phase diagram of $^3$He in nafen-910. Filled circles mark the superfluid transition of $^3$He for 2.5-coverage. Open circles mark the transition for 2.2-coverage. Dashed line is a guide to the eye, but it takes into account that at $P=29.3$\,bar no transition was observed for 2.2-coverage down to $\approx0.3\,T_c$.}
\label{pd910}
\end{figure}

{\it Results in pure $^3$He and for 2.5-coverage.---}In Fig.~\ref{pd72} and Fig.~\ref{mode} superfluid phase diagrams of $^3$He in nafen-72, nafen-90 and nafen-243 for 2.5-coverage and in pure $^3$He are shown. It is seen that in all samples $\Delta T_{ca}$ in pure $^3$He is essentially greater. Moreover, on cooling from the normal phase in presence of the coverage the transition occurs into the polar phase while in pure $^3$He -- into the A phase (at least in nafen-72 and nafen-90 where $\Delta T_{ca}$ is still small). We found that $\Delta T_{ca}$ in pure $^3$He more strongly depends on nafen porosity than in presence of $^4$He. It is especially clear for nafen-910: although its porosity is extremely low, we observed the superfluid transition for 2.5-coverage (Fig.~\ref{pd910}) but in pure $^3$He the transition was not detected at all pressures down to the lowest attained temperatures ($\approx0.25\,T_c$ at 29.3\,bar).

\begin{figure}[t]
\centerline{\includegraphics[width=\columnwidth]{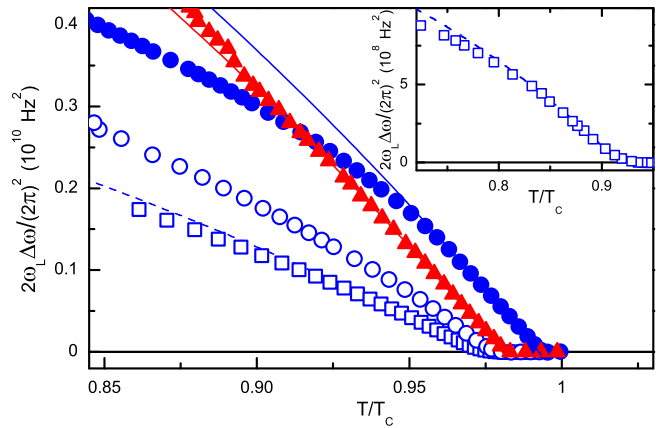}}
\caption{
Cw NMR frequency shifts versus temperature at $P=29.3$\,bar in SN states for $\mu=0$. Filled circles: nafen-72 (2.5-coverage, $T_{ca}=0.993\,T_c$). Triangles: nafen-243 (2.5-coverage, $T_{ca}=0.981\,T_c$). Open circles: nafen-72 (2.2-coverage, $T_{ca}=0.977\,T_c$). Open squares: nafen-72 (pure $^3$He, $T_{ca}=0.974\,T_c$). Solid lines: theory for the polar phase with $K=1.15$. Dashed line: theory for the A phase for $T_{ca}=0.974\,T_c$. (Insert) Open squares: cw NMR frequency shift in pure $^3$He in nafen-72 at $P=7.1$\,bar in the SN state for $\mu=0$. $T_{ca}=0.917\,T_c$. Dashed line: theory for the A phase.}
\label{freq86}
\end{figure}

\begin{figure}[t]
\centerline{\includegraphics[width=\columnwidth]{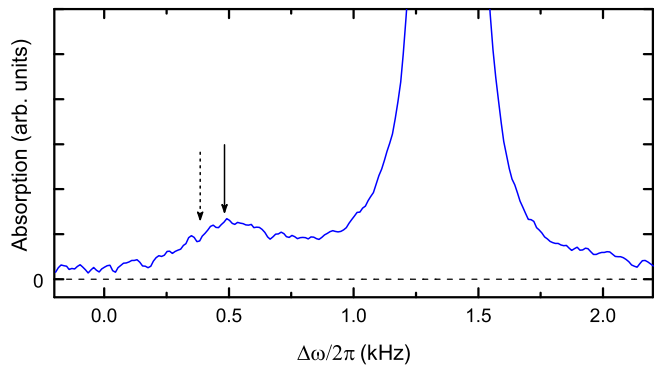}}
\caption{
Cw NMR line with the satellite peak due to pinned half-quantum vortices in $^3$He in nafen-910. Arrows mark the expected position of the satellite peak for $\lambda=0.93$ (solid) and $\lambda=1$ (dashed). $\mu=40^\circ$, $P=23.6$\,bar, $H=16.8$\,mT, $T\approx0.45\,T_c$, $T_{ca}=0.68\,T_c$.}
\label{sat}
\end{figure}

In Fig.~\ref{freq86} by filled symbols we show the temperature dependence of $\Delta\omega$ in $^3$He in nafen-72 and nafen-243 for 2.5-coverage at $P=29.3$\,bar. It is seen that below $T_{ca}$ down to $\approx0.96\,T_c$ the shift in nafen-72 follows the curve with the same slope as for nafen-243 that corresponds to the polar phase. Below $\approx0.96\,T_c$ the shift in nafen-72 deflects from the theoretical curve due to the 2nd order transition into the DA phase, similar to that observed in nafen-90 \cite{we3}. On further cooling, the 1st order transition into a low temperature (LT) phase occurs which NMR properties  correspond to the DB phase. Additionally, our attempts to create the SG state in $^3$He in nafen-72 for 2.5-coverage have failed as it should be if the polar phase exists in some temperature range below $T_{ca}$. In $^3$He in nafen-910 with 2.5-coverage, we observe only one superfluid phase. In this case we cannot compare the measured $\Delta\omega$ with Eq.~(\ref{shift0}) because $\Delta T_{ca}$ for this sample is too large but we are sure that this phase is the polar phase. Firstly, our attempts to create the SG state in this sample have failed. Secondly, for 2.5-coverage we were able to create half-quantum vortices in nafen-910 (as well as in other samples). As it was shown in experiments \cite{Hels}, in the polar phase such vortices can be created by a fast cooling through $T_{ca}$ owing to the Kibble-Zurek mechanism. They survive in nafen due to the pinning and result in a small satellite NMR peak with a frequency shift of $\Delta\omega_{sat}=\Delta\omega_0(\cos^2\mu-\lambda\sin^2\mu)$, where $\Delta\omega_0$ is the shift of the main NMR line for $\mu=0$ and $\lambda$ is close to (but less than) 1. Fig.~\ref{sat} shows the example of the NMR spectrum obtained in nafen-910 after the fast cooling experiment. The satellite peak position corresponds to $\lambda=0.93$ that agrees with Ref.~\cite{Hels}.

\begin{figure}[t]
\centerline{\includegraphics[width=\columnwidth]{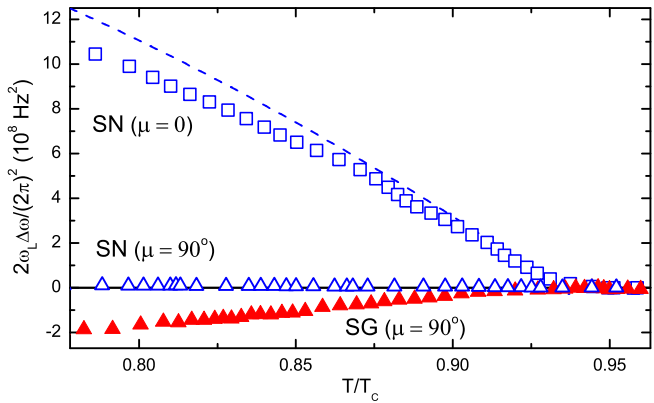}}
\caption{
Cw NMR frequency shifts versus temperature in pure $^3$He in nafen-90. Circles: the SN state for $\mu=0$. Open triangles: the SN state for $\mu=90^\circ$. Filled triangles: the SG state for $\mu=90^\circ$. Dashed line: theory for the A phase for $\mu=0$. $P=12.3$\,bar, $T_{ca}=0.935\,T_c$.}
\label{snsg90}
\end{figure}

In pure $^3$He in aerogel the measured NMR frequency is a weighted average of NMR frequencies of liquid $^3$He and solid paramagnetic $^3$He on the strands due to the fast exchange mechanism \cite{Fr}. A magnetization of solid $^3$He follows the Curie-Weiss law, and at $T\sim T_{ca}$ its total magnetic moment ($M_s$) may exceed the magnetic moment of liquid $^3$He ($M_l$). For example, in nafen-243 at 7.1\,bar $M_s/M_l\approx2.7$ at $T=T_c$. Therefore, to compare experimental data with Eq.~(\ref{shift0}), the observed shift should be corrected that can be done if the temperature dependence of $M_s/M_l$ is measured \cite{Fr}. In Figs.~\ref{freq86} and \ref{snsg90} we show the corrected data for pure $^3$He in nafen-72 and nafen-90 by open squares. They are close to curves expected for the A phase (deflections are probably due to systematic errors in measurements of $M_s/M_l$). Additionally, in pure $^3$He in nafen-72, nafen-90, and nafen-243 the SG state was easily created (filled triangles in Fig.~\ref{snsg90}) that excludes the existence of the polar phase. These observations allow us to affirm that in pure $^3$He in nafen the superfluid transition occurs into the A phase (or into the A phase with a small polar distortion). On further cooling, in nafen-72 and nafen-90 we observe the 1st order transition into the LT phase which is accompanied by a decrease of spin susceptibility and an increase of $\Delta\omega$ for $\mu=0$. In the LT phase $\Delta\omega$ is close to that expected for the B phase, but the susceptibility change was not measured with sufficient accuracy due to a wide NMR line in this state. Therefore, we only assume that the LT phase is close to the B phase.


{\it Results for 2.2-coverage.---}In $^3$He in nafen-910 for 2.2-coverage we detect no signal from the paramagnetic $^3$He at all pressures except 29.3\,bar. At this pressure it is small (corresponds to $\lesssim0.1$ atomic layer of solid $^3$He) and the superfluid transition was not found down to the lowest attained temperature ($\approx0.3\,T_c$). At lower pressures the transition was observed (open circles in Fig.~\ref{pd910}) but $T_{ca}$ is suppressed more than for 2.5-coverage.
Such unusual pressure dependence of $T_{ca}$ is explained by the following. It is known that the amount of $^4$He which is enough to remove solid $^3$He from the surface decreases with the decrease of pressure \cite{K93}.
Therefore, at low pressures 2.2-coverage should be enough to completely remove the paramagnetic $^3$He. However, above $\approx15$\,bar a small amount of solid $^3$He appears and $\Delta T_{ca}$ increases. This amount grows with the increase of pressure, and at 29.3\,bar, where the paramagnetic signal is already detectable, the suppression of $T_{ca}$ becomes so large that we cannot reach $T_{ca}$.

In nafen-72, placed at the same chamber, the paramagnetic signal was not detected even at 29.3\,bar due to higher porosity of the sample. However, the effect of solid $^3$He is also clear: the temperature dependence of $\Delta\omega$ for $\mu=0$ (open circles in Fig.~\ref{freq86}) no longer follows the curve for the polar phase, but corresponds to the DA phase: $K\approx0.75$ near $T_{ca}$ and $K\approx0.66$ at $T=0.85\,T_c$. The influence of solid $^3$He in nafen-72 is seen down to $\approx15$\,bar, but at lower pressures we see no difference between 2.2 and 2.5-coverages. It is worthy to mention that together with these experiments we investigated a behavior of the quartz tuning fork. It was found that the fork resonant properties are also sensitive to the presence of solid $^3$He \cite{qtf}.

{\it Conclusions.---}We observe that even small amount of paramagnetic solid $^3$He on nafen strands drastically changes $^3$He superfluid phase diagram: on cooling from the normal phase the superfluid transition occurs into DA or pure A phases while in the absence of the solid $^3$He it occurs into the polar phase. Solid $^3$He on the strands also essentially reduces $T_{ca}$, especially in low porosity nafen, where the scattering anisotropy is greater \cite{diff0,diff}. The observed phenomena cannot be explained by a change of the scattering specularity because they are observed also at high pressures where the scattering should be diffusive regardless of presence or absence of solid $^3$He. The only explanation which we can suggest is a great importance of the magnetic channel in case of strong anisotropy of the scattering of $^3$He quaiparticles. We hope that our results will serve as a stimulus for further theoretical studies of the effect of magnetic scattering on triplet superfluidity.

\begin{acknowledgments}
We are grateful to I.M.~Grodnensky for providing nafen samples, V.P.~Mineev, J.A.~Sauls and G.E.~Volovik for useful discussions and comments. This work was supported in part by RFBR (Grant No. 16-02-00349) and by Basic Research Program of the Presidium of Russian Academy of Sciences.
\end{acknowledgments}

\end{document}